\documentclass{osa-article}
\journal{osajournal}
\articletype{Research Article}

\usepackage{mathtools, mathrsfs}
\usepackage{caption}
\usepackage{subcaption}

\begin{document}

\title{Optimal Physical Preprocessing for Example-Based Super-Resolution}

\author{Alexander Robey and Vidya Ganapati\authormark{*}}

\address{Swarthmore College, Engineering Department, 500 College Ave, Swarthmore, PA 19081}

\email{\authormark{*}vganapa1@swarthmore.edu} 

\homepage{ganapati.swarthmore.edu}

\begin{abstract}
In example-based super-resolution, the function relating low-resolution images to their high-resolution counterparts is learned from a given dataset. This data-driven approach to solving the inverse problem of increasing image resolution has been implemented with deep learning algorithms. In this work, we explore modifying the imaging hardware in order to collect more informative low-resolution images for better ultimate high-resolution image reconstruction. We show that this ``physical preprocessing" allows for improved image reconstruction with deep learning in Fourier ptychographic microscopy. 

Fourier ptychographic microscopy is a technique allowing for both high resolution and high field-of-view at the cost of temporal resolution. In Fourier ptychographic microscopy, variable illumination patterns are used to collect multiple low-resolution images. These low-resolution images are then computationally combined to create an image with resolution exceeding that of any single image from the microscope. We use deep learning to jointly optimize the illumination pattern with the post-processing reconstruction algorithm for a given sample type, allowing for single-shot imaging with both high resolution and high field-of-view. We demonstrate, with simulated data, that the joint optimization yields improved image reconstruction as compared with sole optimization of the post-processing reconstruction algorithm.
\end{abstract}

\section{Introduction}

Deep learning, a subset of machine learning, has shown remarkable results in the interpretation of data. By computationally finding complex, higher-order patterns in labeled datasets, deep learning has yielded state-of-the-art results in areas such as the classification of images \cite{he_deep_2016}, language translation \cite{sutskever_sequence_2014}, and speech recognition \cite{hinton_deep_2012}. Deep learning is powerful because the mathematical form of the relationship between the input and output need not to be specified beforehand; the deep learning algorithm can computationally find arbitrary, non-linear relationships in datasets.

The ``super-resolution" problem of converting low-resolution images to higher resolution has been attempted with machine learning \cite{freeman_example-based_2002, kim_deeply-recursive_2015, romano_raisr_2016, dong_image_2016, Kim_2016_CVPR, ledig_photo-realistic_2016, sajjadi_enhancenet_2016, dahl_pixel_2017, hayat_super-resolution_2017}. The super-resolution problem is an ill-posed inverse problem, with many high-resolution images possible for a given low-resolution image. In example-based super-resolution, machine learning attempts to make the problem well-posed by applying prior information.

In example-based super-resolution, the training inputs are low-resolution image patches, and the training outputs are the corresponding high-resolution image patches. Machine learning then attempts to find the function that takes the low-resolution patch as the input, and then outputs the high-resolution patch. Deep convolutional neural networks have been promising in tackling the super-resolution problem \cite{kim_deeply-recursive_2015, dong_image_2016, Kim_2016_CVPR, sajjadi_enhancenet_2016, dahl_pixel_2017}, and the addition of adversarial networks has resulted in improved image quality of fine details \cite{ledig_photo-realistic_2016}. Example-based super-resolution with deep learning has also been applied to microscopy images \cite{rivenson_deep_2017, rivenson_deep_2017-1, rivenson_toward_2018}. The success of these super-resolution deep neural networks can be attributed to the fact that images are not random collections of pixels; images contain some structure. The more we constrain the category or type of images we use, the more structure the deep learning algorithm should be able to find.

The desirability of obtaining a high-resolution image from low resolution stems from the trade-offs inherent in designing imaging systems, due to constraints in physics and cost. The minimum distance allowing two points to be distinguishable from each other determines the resolution of an image; the lower the minimum distance, the higher the resolution. In the design of an imaging system, resolution must be traded off with the total area imaged, the field-of-view. We can obtain a low-resolution image with high field-of-view or vice versa. The tradeoff between resolution and field-of-view of an imaging system can be quantified by its space-bandwidth product \cite{goodman_introduction_2005}. Example-based super-resolution is a pathway to obtaining high resolution and high field-of-view at the same time, i.e. a high space-bandwidth product.

Another way to obtain a high space-bandwidth product is to trade off temporal resolution by taking multiple images and stitching them together. This approach is undertaken in some computational imaging modalities, where multiple image measurements are collected and algorithmically combined. 

In the computational imaging paradigm, the imaging hardware is co-designed with the post-processing software. In computational imaging, we do not attempt to collect an image directly. Instead, we collect measurements that are used to create the image in post-processing. The individual measurements do not necessarily resemble the final image, but they contain the information necessary for image reconstruction. The data collected by the image sensor may be a garbled version of the desired final image, requiring considerable computational post-processing, sometimes with an iterative algorithm. The image reconstruction process from the collected noisy sensor data is an inverse problem and may be ill-posed. Computational imaging modalities include fluorescence imaging such as stochastic optical reconstruction microscopy \cite{rust_sub-diffraction-limit_2006}, photoactivated localization microscopy \cite{betzig_imaging_2006, hess_ultra-high_2006} and structured light microscopy \cite{gustafsson_surpassing_2000, gustafsson_nonlinear_2005}, where multiple lower-resolution images are computationally combined for a higher-resolution image without loss of field-of-view. In computational imaging, we can also consider the problem of imaging the phase of light, which cannot be measured directly. Fourier ptychographic microscopy is a modality where multiple lower-resolution images are computationally combined for a higher-resolution reconstruction of a complex object with both phase and amplitude \cite{zheng_wide-field_2013}.

In Fourier ptychographic microscopy, the illumination source is a 2-dimensional matrix of light-emitting diodes (LEDs). Low-resolution images with high field-of-view are collected by illuminating the sample with different patterns of LEDs. The different illumination patterns cause different spatial frequencies to be modulated into the pass-band of the optical system. These low-resolution images are then computationally combined to create a high-resolution reconstruction of the phase and amplitude of the sample, preserving the field-of-view of the low-resolution images. An iterative algorithm is generally used for the reconstruction, but can be computationally intensive. Recently, deep learning has been used to solve the inverse problem in several computational imaging modalities, replacing iterative methods \cite{kamilov_learning_2015, kalantari_learning-based_2016, sinha_solving_2016, li_imaging_2017, mardani_recurrent_2017, chen_low-dose_2017, gul_spatial_2017, jin_deep_2017, shimobaba_computational_2017, sinha_lensless_2017, mccann_review_2017, borhani_learning_2018,  goy_low_2018, nguyen_2d_2018, sun_efficient_2018, wu_extended_2018-1}, including in Fourier ptychographic microscopy \cite{nguyen_convolutional_2018}.

Though Fourier ptychographic microscopy achieves a high space-bandwidth product, temporal resolution is sacrificed: many low-resolution images need to be collected for reconstruction. This makes imaging of dynamic processes, such as those in live cells, difficult. Imaging of such processes requires sophisticated hardware control to rapidly acquire the low-resolution image stack \cite{tian_computational_2015}. 

An emerging application for deep learning is to learn how to both optimally collect and process data for some end goal, rather than just process data \cite{adler_compressed_2016, chakrabarti_learning_2016, horstmeyer_convolutional_2017, lohit_convolutional_2017, haim_deep_2018}. With this ``physical preprocessing" procedure for deep learning, data acquisition time or hardware requirements can be reduced. In this work, we use deep learning to jointly optimize the illumination pattern in Fourier ptychographic microscopy with the reconstruction algorithm for a given sample type, allowing for single-shot imaging with a high space-bandwidth product. Our deep learning algorithm looks for the illumination pattern that will embed the most information into a single low-resolution image, along with a direct reconstruction method to non-iteratively reconstruct the high-resolution complex field from the single image measurement. 

\section{Background on Fourier Ptychographic Microscopy}

\begin{figure}[htbp]
    \centering
    \includegraphics[width=7cm]{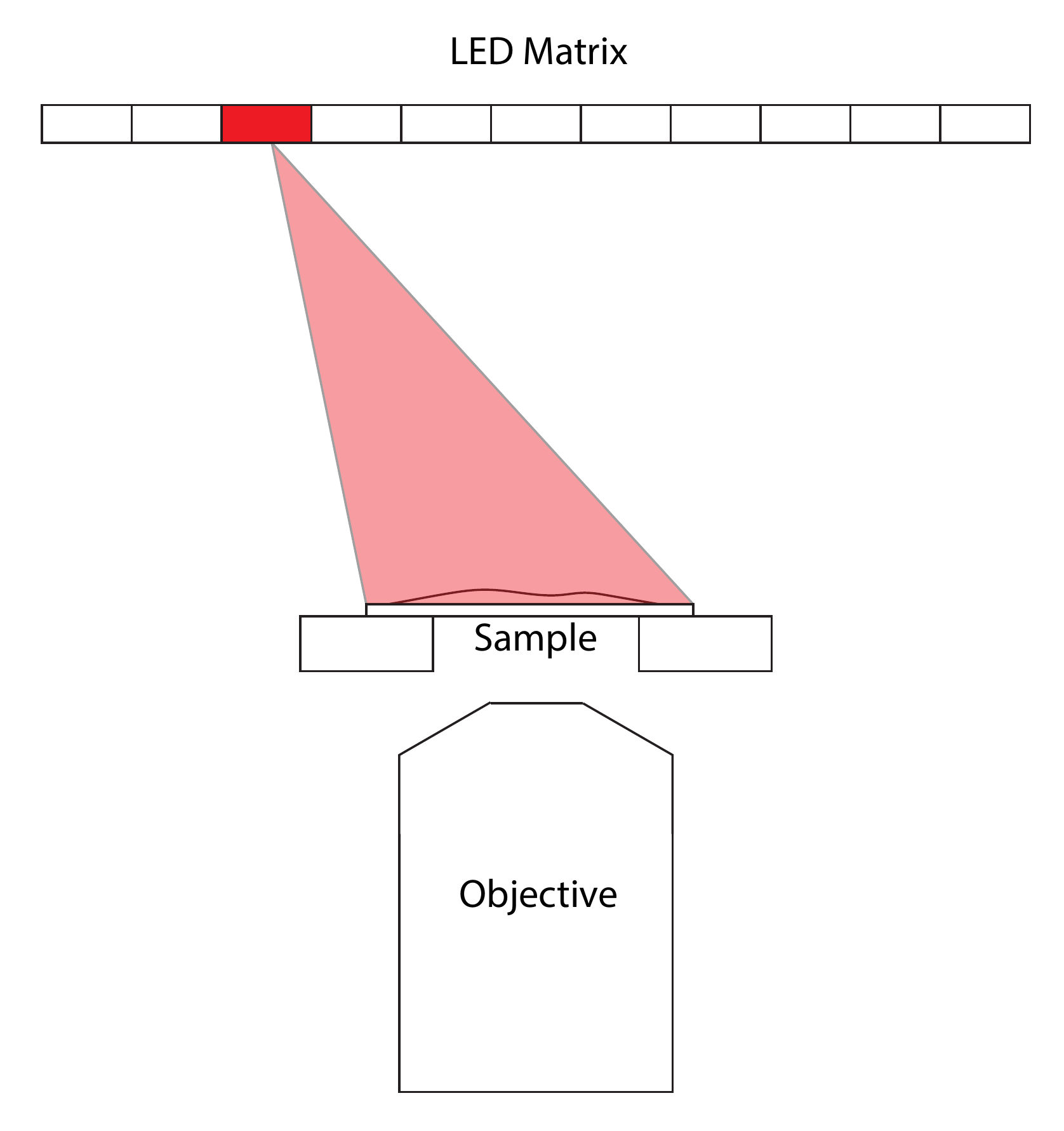}
    \caption{Schematic of Fourier ptychographic microscope setup. An LED matrix replaces the illumination source of a traditional microscope.}
    \label{fig:schematic}
\end{figure}

In Fourier ptychographic microscopy, a sample is illuminated by a light-emitting diode (LED), as in Fig.~\ref{fig:schematic}. We assume that the LED is sufficiently far away that the light entering the sample can be approximated as a plane wave. The sample scatters the entering plane wave, and the exiting wave is imaged through a microscope objective with a given numerical aperture (NA). 

\begin{figure}[htbp]
    \centering
    \includegraphics[width=9cm]{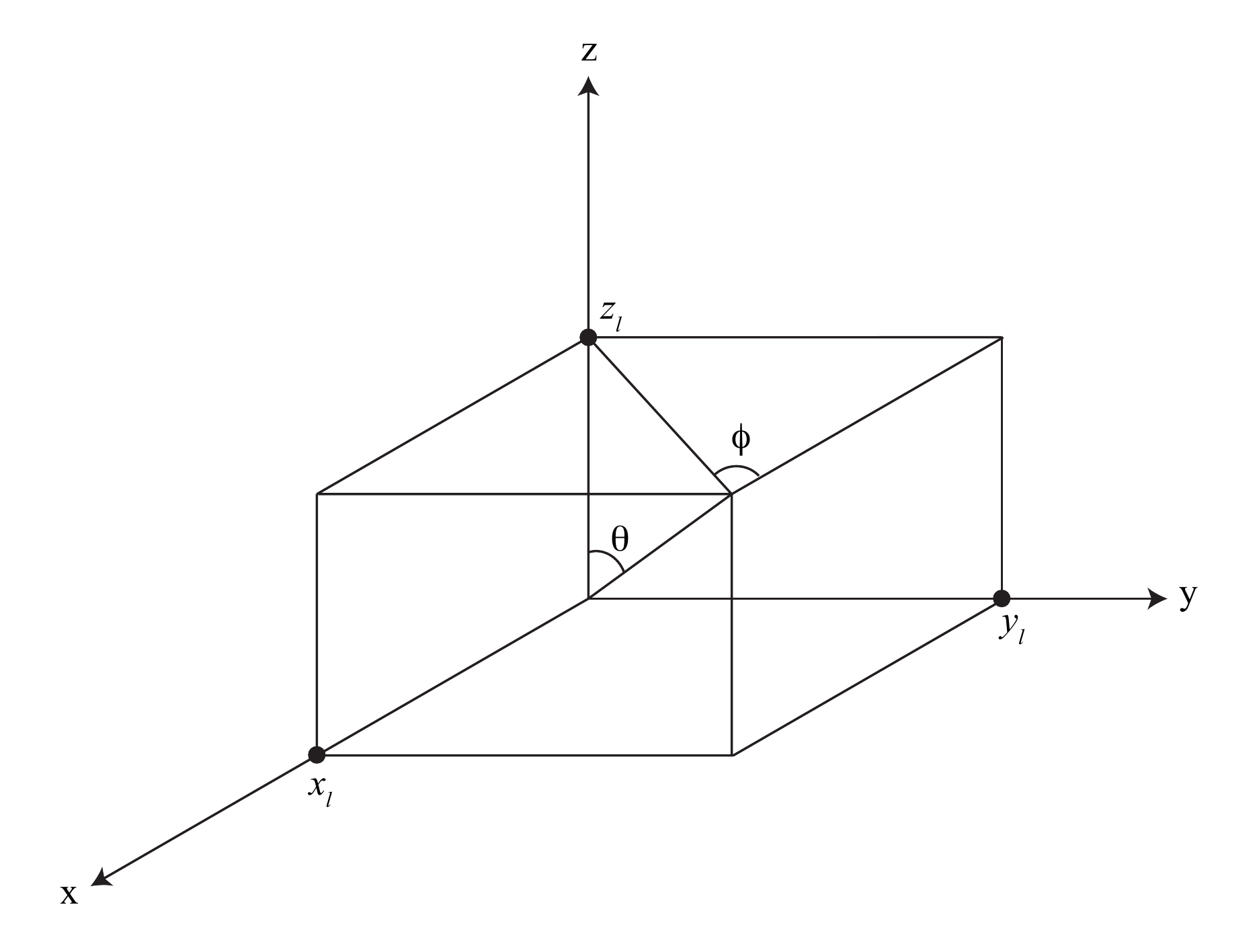}
    \caption{The light from an LED at point $\left(  x_{l}, y_{l}, z_{l} \right)$ illuminating a sample centered at the origin can be approximated as a plane wave with spatial frequencies given by Eqn.~\ref{eq:ulx} and Eqn.~\ref{eq:uly}.}
    \label{fig:graph}
\end{figure}

Consider a thin sample that is in the $xy$-plane and centered at the origin in in the Cartesian coordinate system shown in Fig.~\ref{fig:graph}. The illuminating LED is placed at the point $ \left(  x_{l}, y_{l}, z_{l} \right)$. The spatial profile of the plane wave from the LED entering the sample is described by $e^{i 2 \pi \vec{u_l} \cdot \vec{r}}$ where $ \vec{u_l} \cdot \hat{x} = u_{l,x} $ is the spatial frequency in the $x$-direction and $ \vec{u_l} \cdot \hat{y} = u_{l,y} $ is the spatial frequency in the $y$-direction. The vector $\vec{u_l}$ is in the direction of $-x_{l} \hat{x} -y_{l} \hat{y} - z_{l} \hat{z} $. The magnitude of $\vec{u_l}$ is $\mid\vec{u_l}\mid = \frac{1}{\lambda}$, where $\lambda$ is the wavelength of the light. From Fig.~\ref{fig:graph}, we see the following:

\begin{equation}
u_{l,x} = - \frac{1}{\lambda} \sin \theta \cos \phi  = -\frac{x_{l}}{\lambda \sqrt{x_{l}^2+y_{l}^2+z_{l}^2}},
\label{eq:ulx}
\end{equation}

\begin{equation}
u_{l,y} = -  \frac{1}{\lambda} \sin \theta \sin \phi  = -\frac{y_{l}}{\lambda \sqrt{x_{l}^2+y_{l}^2+z_{l}^2}}.
\label{eq:uly}
\end{equation}

We use the thin transparency approximation to describe the sample as a complex transmission function $o(x,y)$. The field immediately after the sample can be described as the object transmission function multiplied by the illumination plane wave at $z=0$:
\begin{equation} 
o(x,y) e^{ i 2 \pi \left(u_{l,x} x + u_{l,y} y \right)}.
\label{eq:exit-wave}
\end{equation}

If the 2D Fourier transform of $o(x,y)$ is given as $O(u_x, u_y)$, the field in Eqn.~\ref{eq:exit-wave} in Fourier space is simply a shift, $O(u_x - u_{l,x}, u_y - u_{l,y})$. The field is multiplied in Fourier space by the pupil function of the objective, $P(u_x, u_y)$, a low-pass filter. In the case of no aberrations, $P(u_x, u_y)$ is unity within a circle with radius $\frac{\text{NA}}{\lambda}$, and zero elsewhere. The image sensor records the intensity $I$ of this field in real space:

\begin{equation}
I = \mid \mathscr{F}^{-1} \{ P(u_x, u_y)O(u_x - u_{l,x}, u_y - u_{l,y}) \} \mid^2 ,
\end{equation}

\noindent where $\mathscr{F}^{-1}$ denotes the inverse 2D Fourier transform \cite{tian_multiplexed_2014, yeh_experimental_2015}. Due to the low-pass filtering operation, the intensity image contains information from only a portion of spatial frequencies of the original object.

The previous discussion assumes illumination from a single LED, where it is assumed that the LED emits a coherent wave. In an array of LEDs, with multiple LEDs illuminated, we assume that each LED emits a coherent wave that is mutually incoherent with the waves emitted from the other LEDs. Thus, if we have an matrix of LEDs in the $xy$-plane at height $z_l$, the intensity image recorded is:

\begin{equation}
I = \sum \limits_{l=1}^n c_{l} \mid \mathscr{F}^{-1} \{ P(u_x, u_y)O(u_x - u_{l,x}, u_y - u_{l,y}) \} \mid^2 ,
\end{equation}

\noindent where $n$ is the total number of illuminated LEDs and $c_l$ is the intensity of LED $l$ \cite{tian_multiplexed_2014}.

In the original instantiation of Fourier ptychographic microscopy, the LEDs are turned on and off one at a time, generating $n$ intensity images \cite{zheng_wide-field_2013}. Other work uses a series of patterns of illuminated LEDs, reducing the total number of images needed \cite{tian_multiplexed_2014, tian_computational_2015}. Adaptive schemes can also reduce the required number of LED images \cite{zhang_self-learning_2015}.

The goal of Fourier ptychographic microscopy is to reconstruct the sample's complex transmission function, $o(x,y)$ from these intensity images. A variety of iterative approaches have been applied to the reconstruction process \cite{yeh_experimental_2015}. The creation of each intensity image involves frequency modulation and a low-pass filtering operation. Each intensity image has limited spatial bandwidth, but reflects information from a different portion of frequency space. Thus, the reconstructed $o(x,y)$ in Fourier ptychographic microscopy can have higher spatial bandwidth than that specified by the NA of the objective. 

Though Fourier ptychographic microscopy creates images with a high space-bandwidth product, it requires acquisition of multiple images. In this work, we aim to enable Fourier ptychographic microscopy with single image acquisition, using a deep learning approach to jointly find the optimal illumination pattern and reconstruction procedure for a given sample type. Our reconstruction procedure replaces previously used iterative methods with a deep neural network.

\section{A Communications Approach}

\begin{figure}[htbp]
    \centering
    \includegraphics[scale=0.4]{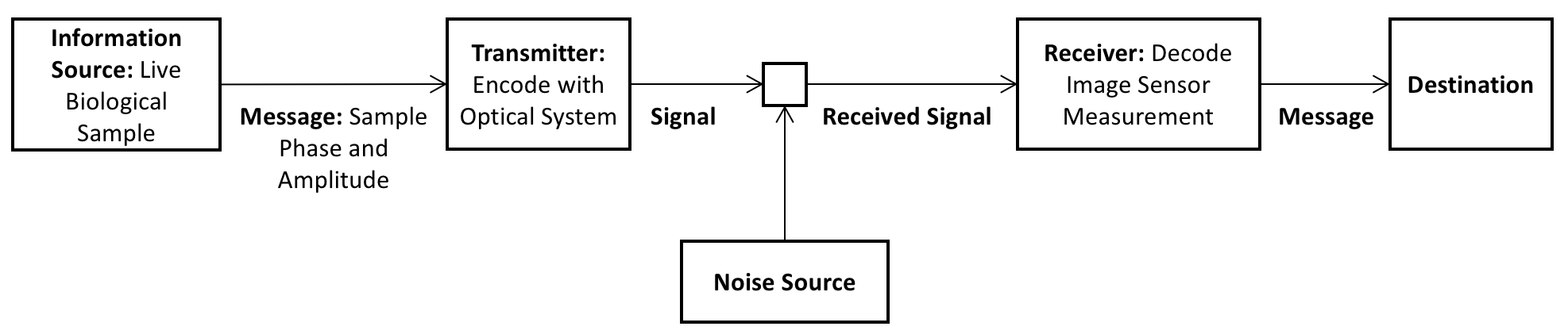}
    \caption{The general communication system in \cite{shannon_mathematical_1948} applied to imaging a thin biological specimen.}
    \label{fig:shannon}
\end{figure}

Claude Shannon's fundamental work in the theory of communication forms a framework to circumvent the loss of temporal resolution in Fourier ptychographic microscopy.  We consider the problem of imaging a live biological specimen and model the Fourier ptychographic microscope as a communication system, as shown in Fig.~\ref{fig:shannon}. In a general communication system, a message from an information source is encoded, transmitted over a noisy channel, and decoded by a receiver. If we know the statistics of the source, and the noise characteristics of the channel, we can optimize the method of encoding and decoding for information transmission \cite{shannon_mathematical_1948}. This communications approach to imaging has been previously explored in \cite{osullivan_information-theoretic_1998}.

In the case of Fourier ptychographic microscopy, a biological sample is the information source, and the message is the sample's phase and amplitude (i.e. its complex transmission function) at a point in time. The message is encoded as an optical signal by the pattern of illuminated LEDs. This optical signal is then transmitted through the optics of the microscope. The image sensor of the microscope camera receives the signal as intensity measurements over an array of sensor pixels. The received signal may be corrupted by noise, such as Poisson noise. Finally, this image sensor measurement must be decoded with an algorithm to try to reconstruct the original phase and amplitude of the sample.

In Fourier ptychographic microscopy, as in most imaging modalities, generally no assumptions are made on the sample being imaged. Except in some cases \cite{yeh_experimental_2015}, the imaging procedure of acquiring low-resolution intensity images and solving for the high-resolution complex object remains invariant with regards to the sample. However, the communication framework in Fig.~\ref{fig:shannon} yields insight on how to reduce the time needed to obtain high space-bandwidth complex objects. If we can determine the statistics of the sample's phase and amplitude ``messages," it might be possible to compress all the information needed for reconstruction of the message into a single image sensor measurement. This physical compression may be achieved by choosing the optimal LED illumination pattern. We think of the sample as having a range of different geometrical configurations, and optimizing a communication system for transmitting and receiving information regarding the current configuration. 

Consider the following simple thought experiment. Imagine a sample having 2 possible states that lead to 2 different phase and amplitude images. With a standard Fourier ptychographic microscope, we would have to take multiple intensity images and complete the iterative reconstruction procedure. However, only 1 bit of information needs to be transmitted and received in this hypothetical system. Intuitively, the standard procedure is wasteful in this simple case; we should easily be able to design an imaging system that only needs a single-shot measurement to find the current state.

There are two major difficulties in the implementation of a communications approach to biological imaging.  First, we need to determine the probability distribution of our ``messages," which we need to discover through observations of the biological sample. Second, we need to encode our messages for transmission, but we cannot choose an arbitrary encoding algorithm. We are constrained to what is physically possible with optics. 

\begin{figure}[htbp]
    \centering
    \includegraphics[scale=0.4]{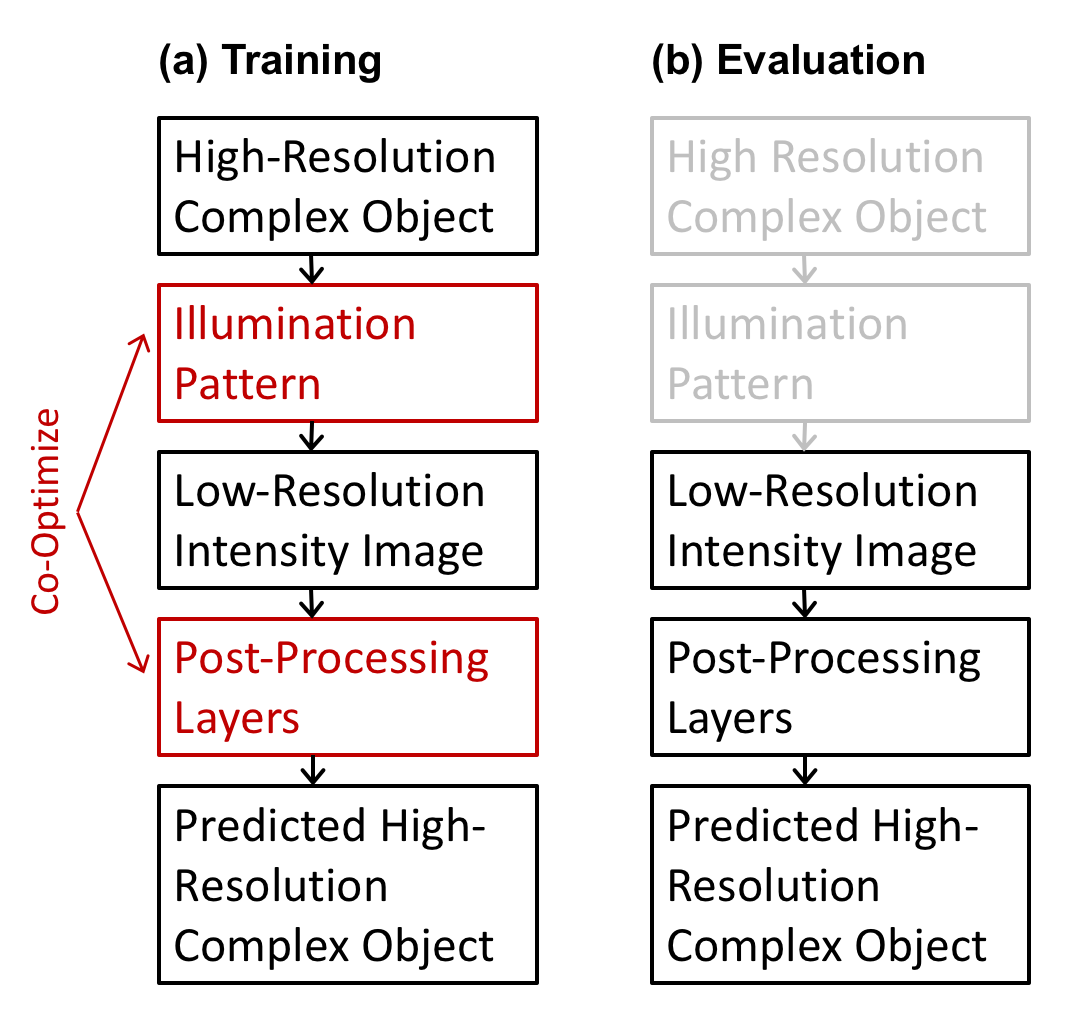}
    \caption{Outline of the training and evaluation of the deep neural network.}
    \label{fig:diagram}
\end{figure}

To solve these problems, we turn to deep learning to optimize the end-to-end communications system, as in \cite{rongwei_joint_2003, oshea_introduction_2017}, simulating the entire end-to-end imaging pipeline as a deep neural network. During training of the network, the hardware parameters for encoding and the software parameters for decoding are optimized. In the training process, the neural network takes high-resolution complex objects obtained from a standard Fourier ptychographic microscope as input. From the input complex object, the low-resolution image collected with a single LED illumination pattern is emulated. This single low-resolution image then passes through post-processing neural network layers, outputting a prediction of the high-resolution complex object. The illumination pattern and the parameters of the post-processing layers are optimized during training, using a dataset of high-resolution complex objects from a given sample type, as diagrammed in Fig.~\ref{fig:diagram}. In the evaluation phase, we implement the optimized illumination in the actual Fourier ptychographic microscope. Each collected single low-resolution image is then fed directly into trained post-processing layers to reconstruct the high-resolution complex object (see Fig.~\ref{fig:diagram}). With this procedure, we could use fixed biological samples to collect the training dataset, and live samples in the evaluation step, potentially enabling unprecedented imaging of dynamic biological processes. 

\section{Methods}

\begin{figure}[htbp]
    \centering
    \includegraphics[scale=0.4]{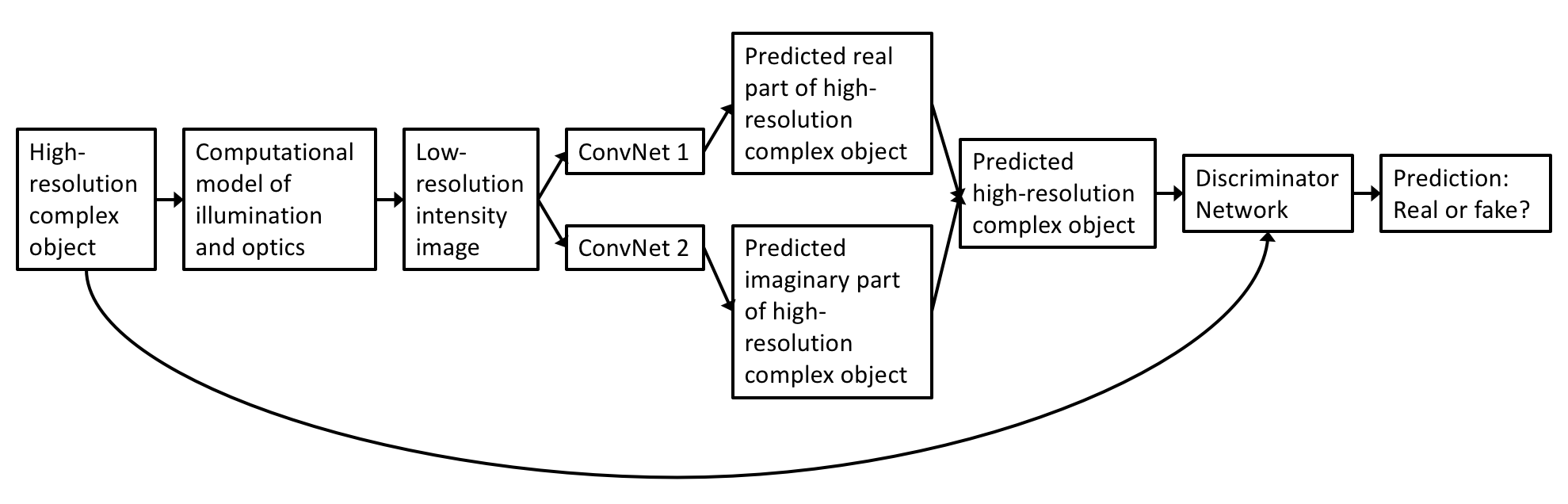}
    \caption{Overview of our computational graph that is trained with a dataset of input examples.}
    \label{fig:network-overview}
\end{figure}

We implement the structure shown in Fig.~\ref{fig:diagram} as a computational graph, and optimize the parameters with TensorFlow \cite{abadi_tensorflow_2016}, an open-source Python package for machine learning. We utilized computational resources from the Extreme Science and Engineering Discovery Environment (XSEDE) \cite{towns_xsede_2014} for training of the computational graph. We include the simulation of the optical function to go from the high-resolution complex object to the low-resolution collected image in the first layers of the deep neural network graph. The latter layers of the deep neural network graph aim to transform the simulated low-resolution image back to the high-resolution complex field, replacing an iterative algorithm. The first ``physical preprocessing" layers emulate the function of the optics, and allow the parameters of the actual physical system to be optimized at the same time as the reconstruction algorithm. An overview of our computational graph is shown in Fig.~\ref{fig:network-overview}.

We utilize computationally simulated datasets for training and testing of the deep neural network. The dataset is composed of high-resolution, high field-of-view complex objects, emulating those reconstructed from a standard Fourier ptychographic microscope setup. To create the complex phase and amplitude images, we take datasets of intensity images and turn them into complex objects. Our first example dataset takes the MNIST dataset of handwritten digits \cite{lecun-mnisthandwrittendigit-2010} and turns it into a complex object dataset by applying the following formula to each pixel:

\begin{equation}
e^{- i \frac{\pi}{2} p_{0} },
\label{eq:mnist}
\end{equation}

\noindent where $p_{0}$ is a pixel of the original image. We then low-pass filter the complex field with the synthetic NA associated with the Fourier ptychographic setup. The resulting complex objects are the inputs to the computational graph in Fig.~\ref{fig:network-overview}.

The low-resolution intensity image corresponding to a single LED illumination pattern, $I_{low}$, is computationally emulated by the procedure outlined in \cite{tian_multiplexed_2014}, summarized in Section 2. We assume a square grid LED matrix where the intensity of the LED can be tuned from 0 to 1, $0 \leq c_l \leq 1$. We emphasize that these physical preprocessing layers are included in the TensorFlow computational graph so that $c_l$ is optimized during the training process. 

The next layers of the computational graph add a Gaussian approximation of Poisson noise to $I_{low}$. Every pixel of $I_{low}$ is processed by:

\begin{equation}
\text{max} \left( \frac{\sqrt{I_{low} \times m} \times g + I_{low} \times m}{m}, 0 \right),
\label{eq:noise}
\end{equation}

\noindent where $m$ is a multiplicative factor chosen to fit the noise of a particular setup and $g$ is drawn from a normal random distribution. A higher $m$ corresponds to higher signal-to-noise ratio. We redraw the value of $g$ at every evaluation of the TensorFlow computational graph.

\begin{figure}[htbp]
    \centering
    \includegraphics[scale=0.4]{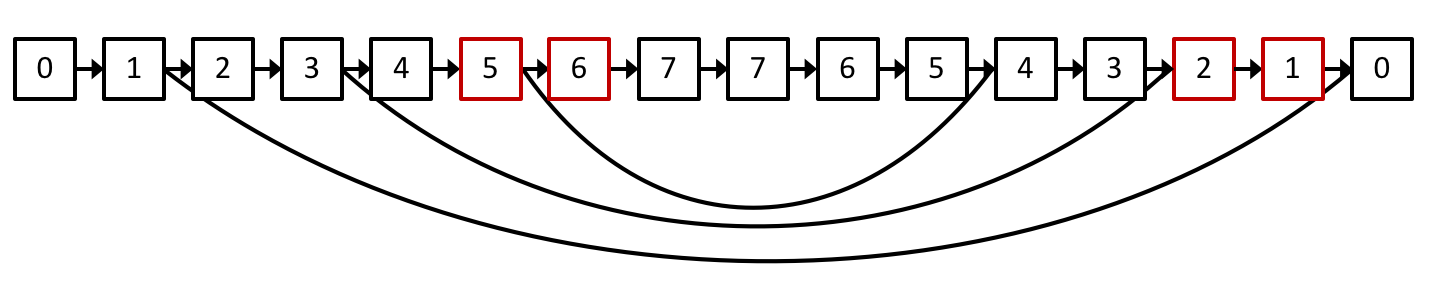}
    \caption{Diagram of the architecture inside the ``ConvNet" blocks in Fig.~\ref{fig:network-overview}. Each numbered square in this diagram represents a convolutional layer with batch normalization and maxout neurons, as shown in Fig.~\ref{fig:ConvNetLayer}. The numbered squares correspond to kernel lengths of 10, 20, 30, 40, 50, 60, 70, 80 respectively. During training, 20\% of the nodes in the layers in red are dropped out. Also shown are the residual connections.}
    \label{fig:ConvNet}
\end{figure}

\begin{figure}[htbp]
    \centering
    \includegraphics[scale=0.4]{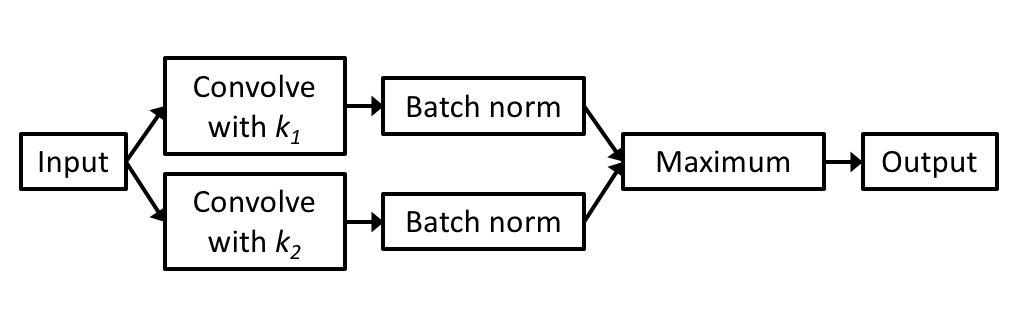}
    \caption{Diagram of each convolution layer in Fig.~\ref{fig:ConvNet}.}
    \label{fig:ConvNetLayer}
\end{figure}

The noisy $I_{low}$ is the input to 2 separate post-processing networks. One network aims to reproduce the real part of the high-resolution object and the other network aims to reproduce the imaginary part, as shown in Fig.~\ref{fig:network-overview}. The architecture for the networks is inspired by \cite{mao_image_2016}, and shown in Fig.~\ref{fig:ConvNet}. Each convolutional layer in Fig.~\ref{fig:ConvNet} is diagrammed in Fig.~\ref{fig:ConvNetLayer}. The networks make use of convolutional layers \cite{lecun_object_1999}, maxout neurons \cite{pmlr-v28-goodfellow13}, batch normalization \cite{ioffe_batch_2015}, residual layers \cite{he_deep_2016}, and dropout \cite{JMLR:v15:srivastava14a}. The outputs of the two neural networks, $I_{high_{real}}$ and $I_{high_{imag}}$, are combined to output the high-resolution complex field reconstruction, $\widetilde{I}_{high} = I_{high_{real}}+ i I_{high_{imag}}$.

Generative adversarial networks have had success in mimicking the probability distribution of a dataset \cite{goodfellow_generative_2014}, and the inclusion of a discriminative network has shown success in improving image quality in super-resolution deep neural networks \cite{ledig_photo-realistic_2016}. We include a discriminative network that takes as input both $\widetilde{I}_{high}$ and the actual high-resolution field $\widetilde{I}_{actual}$ and classifies whether the input is an output of the reconstruction network (a ``fake") or an actual high-resolution image, as shown in Fig.~\ref{fig:network-overview}. The objective function of this discriminative network is the cross-entropy loss of correct classification. We alternate training the parameters of the discriminative network to improve the cross-entropy loss with training of the rest of the computational graph to improve the overall objective function.

Our overall objective function consists of 3 parts, as in \cite{mathieu_deep_2015}:

\begin{equation}
J = M + \alpha G + C,
\end{equation}

\noindent where $M$ is the mean-squared error between $\widetilde{I}_{high}$ and $\widetilde{I}_{actual}$,  $G$ is the mean-squared error between the first gradients of $\widetilde{I}_{high}$ and $\widetilde{I}_{actual}$, and $C$ is the cross-entropy loss representing how well the discriminative network is ``fooled." In calculating $G$, we take the sum of the mean-squared error of the gradient in the vertical direction and the gradient in the horizontal direction. The gradient is calculated by taking the difference  between $\widetilde{I}$ and $\widetilde{I}$ shifted by one place. The metric $C$ measures how well the discriminative network is fooled, by calculating the cross-entropy loss of incorrect classification of $\widetilde{I}_{high}$. The objective function includes the weighting factor $\alpha = 1,000$.

The parameters of the computational graph are trained with the Adam optimizer \cite{kingma_adam_2014}. A randomly selected batch of 4 high-resolution inputs $\widetilde{I}_{actual}$ from the training dataset are used in every iteration of training. For each example in this work, we train for 100,000 iterations, with a training rate of $1 \times 10^{-2}$. The training rate is decayed by a factor of 0.99 every 1,000 iterations. The variable parameters of the computational graph are initialized with a truncated normal distribution with $\sigma = 0.1$, and values limited to between 2 standard deviations from $\mu = 0$. An exponential running average with decay rate of 0.999 of the variable parameters is saved for testing.

\section{Results and Discussion}


We obtain image reconstruction results when we optimize the entire computational graph in Fig.~\ref{fig:network-overview}. We compare to the results when only optimizing the post-processing layers, keeping the LED illumination pattern constant. As described in the previous section, the MNIST dataset of intensity images is processed by Eqn.~\ref{eq:mnist} and then low-pass filtered with the synthetic NA of the Fourier ptychographic microscope. The optical parameters of the emulated Fourier ptychographic microscope are given in Table~\ref{table:MNIST_Distinct}. We utilize the full MNIST dataset, consisting of 55,000 training images, 5,000 validation images, and 10,000 test images.

\begin{table}
\small
 \begin{center}
  \begin{tabular}{ | l | l |  }
    \hline
    Size of high-resolution object & 5.2 $\mu$m, 32 $\times$ 32 pixels \\ \hline 
    Wavelength of LED illumination & 630 nm \\ \hline 
    NA of microscope objective & 0.3 \\ \hline 
    Magnification of microscope objective & 10$\times$ \\ \hline 
    Pixel size of image sensor & 6.5 $\mu$m \\ \hline 
    Pitch of LED matrix & 4 mm \\ \hline 
    LED matrix size & 7 x 7 grid, center 45 LEDs allowed to vary \\ \hline 
    Height of LED matrix, $z_l$ & 25 mm \\ \hline 
    Synthetic NA of Fourier ptychographic microscope & 0.80 \\ \hline 
  \end{tabular}
\end{center}
\caption{Optical parameters used for the simulated MNIST dataset; the original MNIST dataset is padded with zeros to create 32 $\times$ 32 pixel images. \label{table:MNIST_Distinct}}
\end{table}

We train the following 4 cases of deep neural networks: 
\begin{enumerate}
\item The LED matrix is \emph{fixed} at constant, uniform illumination of $c_l = 1$ for all $l$ during training. 
\item The LED matrix is \emph{initialized} at constant, uniform illumination of $c_l = 1$ for all $l$, but allowed to change during training.
\item The LED matrix is \emph{fixed} at a random initialization of $c_l$, picked from a uniform random distribution.
\item the LED matrix is \emph{initialized} at the same random initialization of $c_l$ as in case (3), but is allowed to vary during training.
\end{enumerate}

\begin{figure}[htbp]
    \centering
    \includegraphics[scale=0.4]{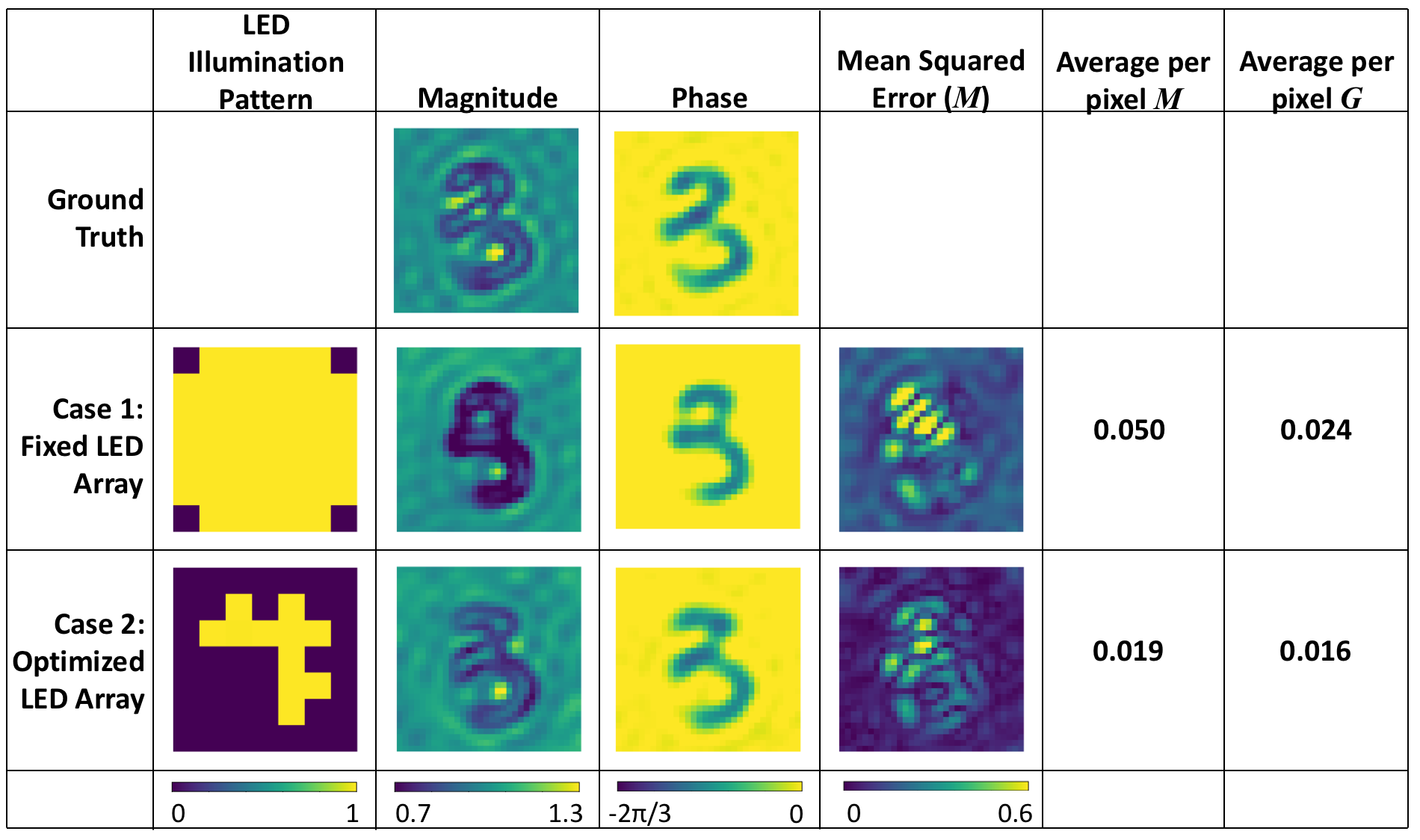}
    \caption{Test results for a single example for cases (1) and (2). The noise factor $m = 0.25$.}
    \label{fig:MNIST_Distinct_ioeo_chart}
\end{figure}

\begin{figure}
\centering
\begin{subfigure}{.5\textwidth}
  \centering
  \includegraphics[scale=0.25]{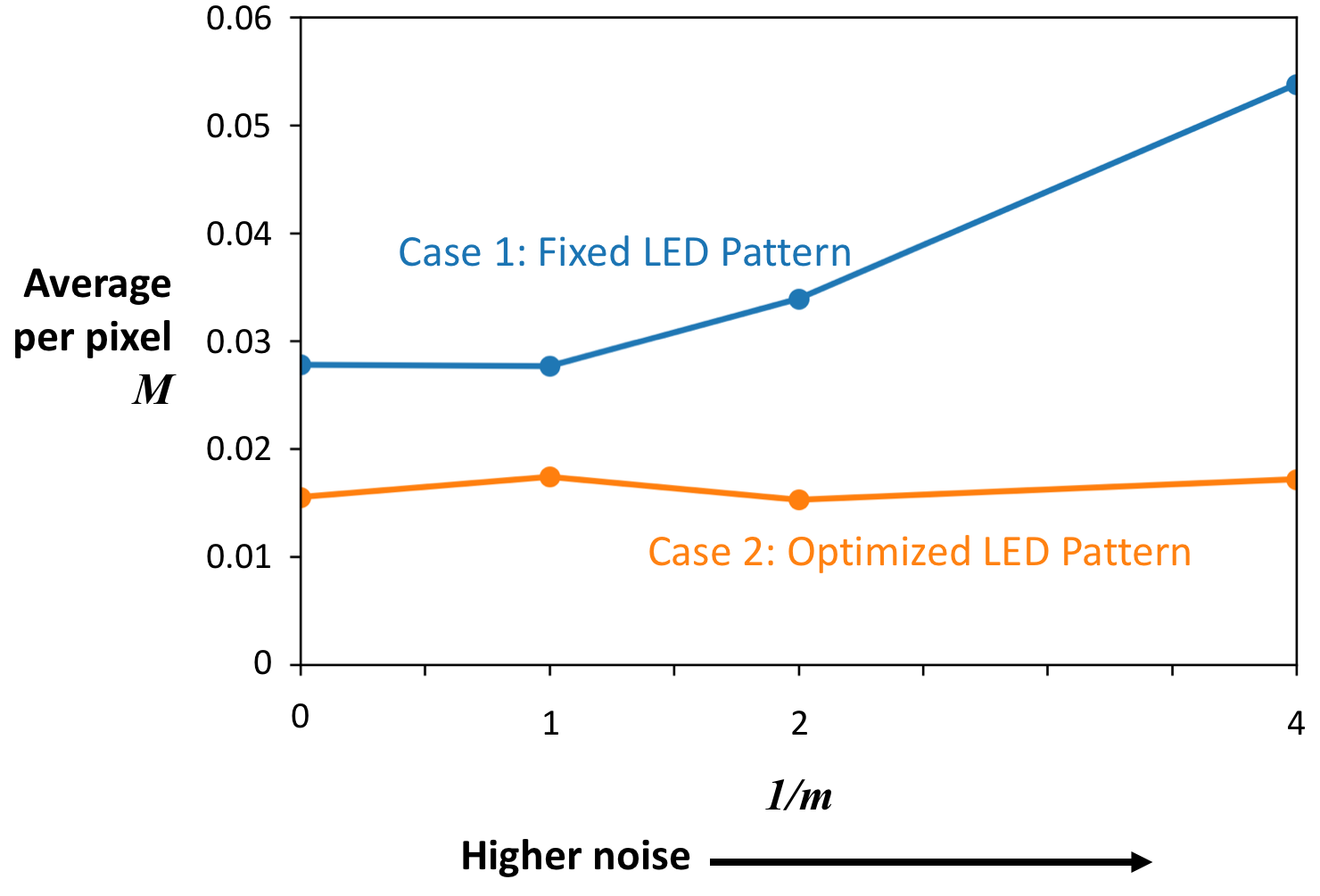}
  \caption{$M$ averaged over the test dataset for different $m$.}
  \label{fig:MNIST_Distinct_ioeo_M_graph}
\end{subfigure}%
\begin{subfigure}{.5\textwidth}
  \centering
  \includegraphics[scale=0.25]{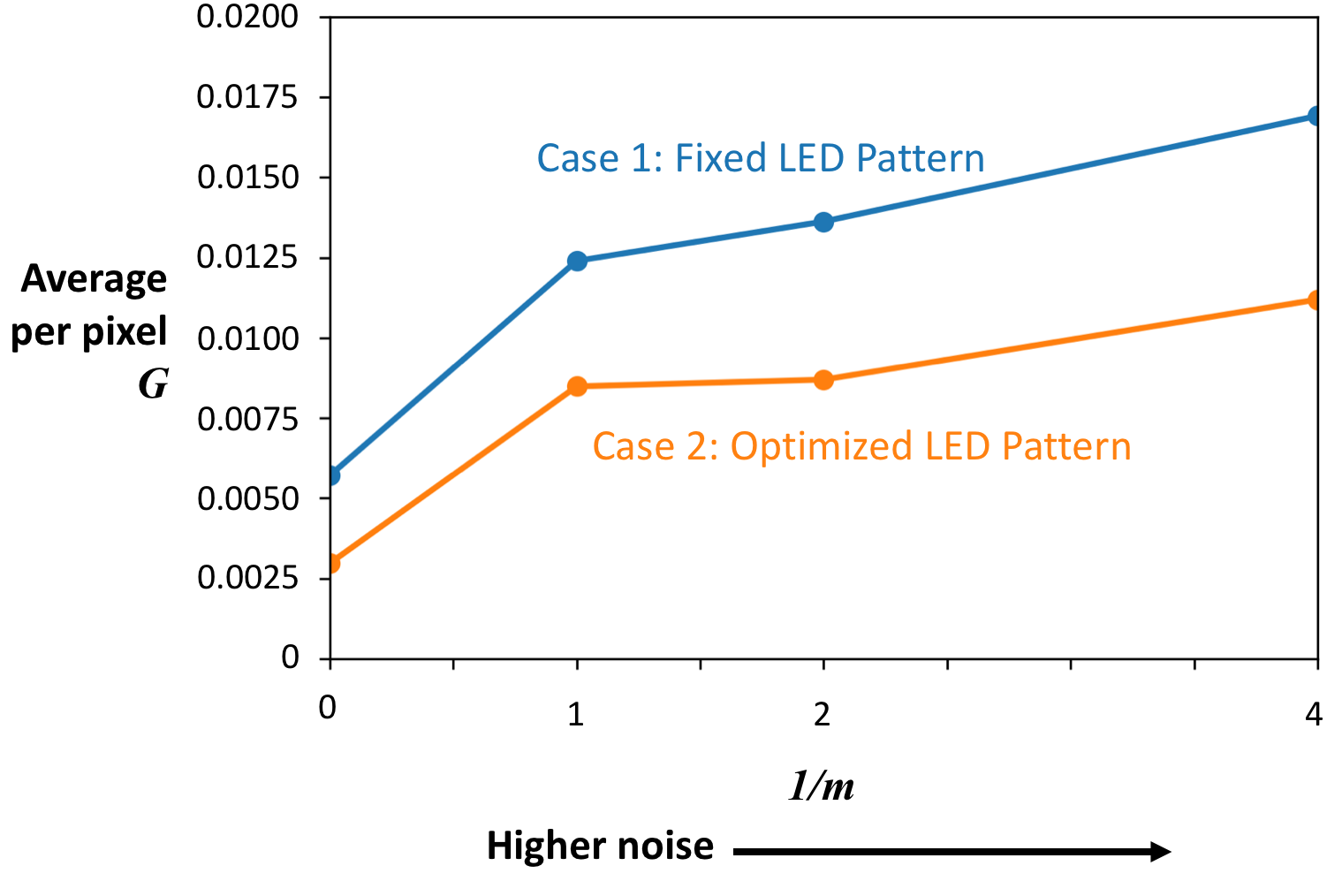}
  \caption{$G$ averaged over the test dataset for different $m$.}
  \label{fig:MNIST_Distinct_ioeo_G_graph}
\end{subfigure}
\caption{Cases (1) and (2).}
\label{fig:MNIST_Distinct_ioeo_graph}
\end{figure}

Fig.~\ref{fig:MNIST_Distinct_ioeo_chart} shows the results of the trained network for a randomly selected test example for cases (1) and (2). In Fig.~\ref{fig:MNIST_Distinct_ioeo_chart}, the noise factor $m = 0.25$. From this single test example, we see that by allowing the illumination pattern to vary, we obtain better results for both $M$, the mean-squared error, and $G$, the mean-squared error of the image gradient. In Fig.~\ref{fig:MNIST_Distinct_ioeo_graph}, we vary $m$, and calculate $M$ and $G$ averaged over the entire test dataset.  Allowing the illumination pattern to vary yields lower $M$ and $G$ for all the noise levels, with the difference growing greater for higher values of $\frac{1}{m}$, corresponding to higher levels of noise. It appears that physical encoding of the data by the illumination pattern allows for more robustness to noise.

\begin{figure}[htbp]
    \centering
    \includegraphics[scale=0.4]{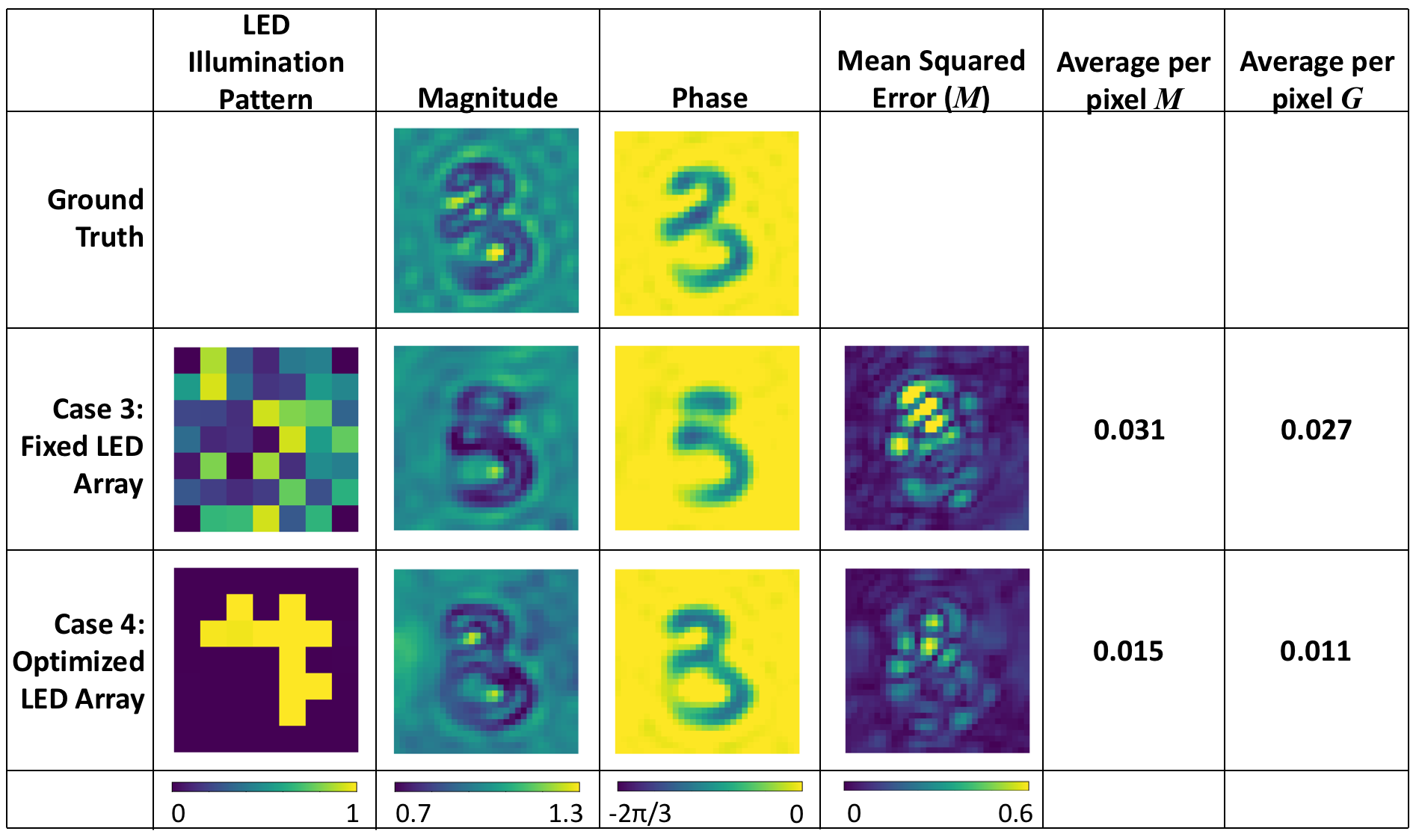}
    \caption{Test results for a single example for cases (3) and (4). The noise factor $m = 0.25$.}
    \label{fig:MNIST_Distinct_none_chart}
\end{figure}

\begin{figure}
\centering
\begin{subfigure}{.5\textwidth}
  \centering
  \includegraphics[scale=0.25]{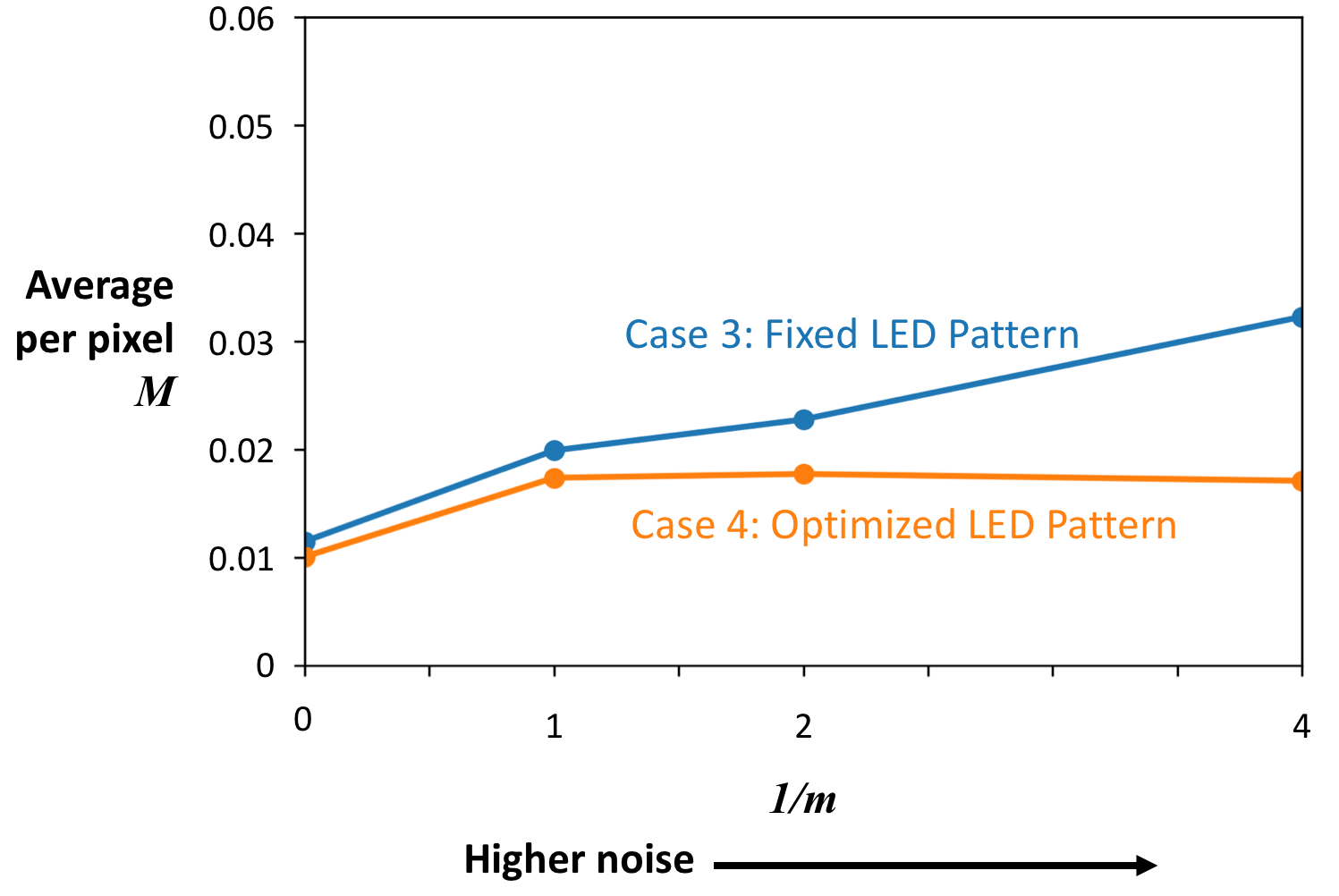}
  \caption{$M$ averaged over the test dataset for different $m$.}
  \label{fig:MNIST_Distinct_none_M_graph}
\end{subfigure}%
\begin{subfigure}{.5\textwidth}
  \centering
  \includegraphics[scale=0.25]{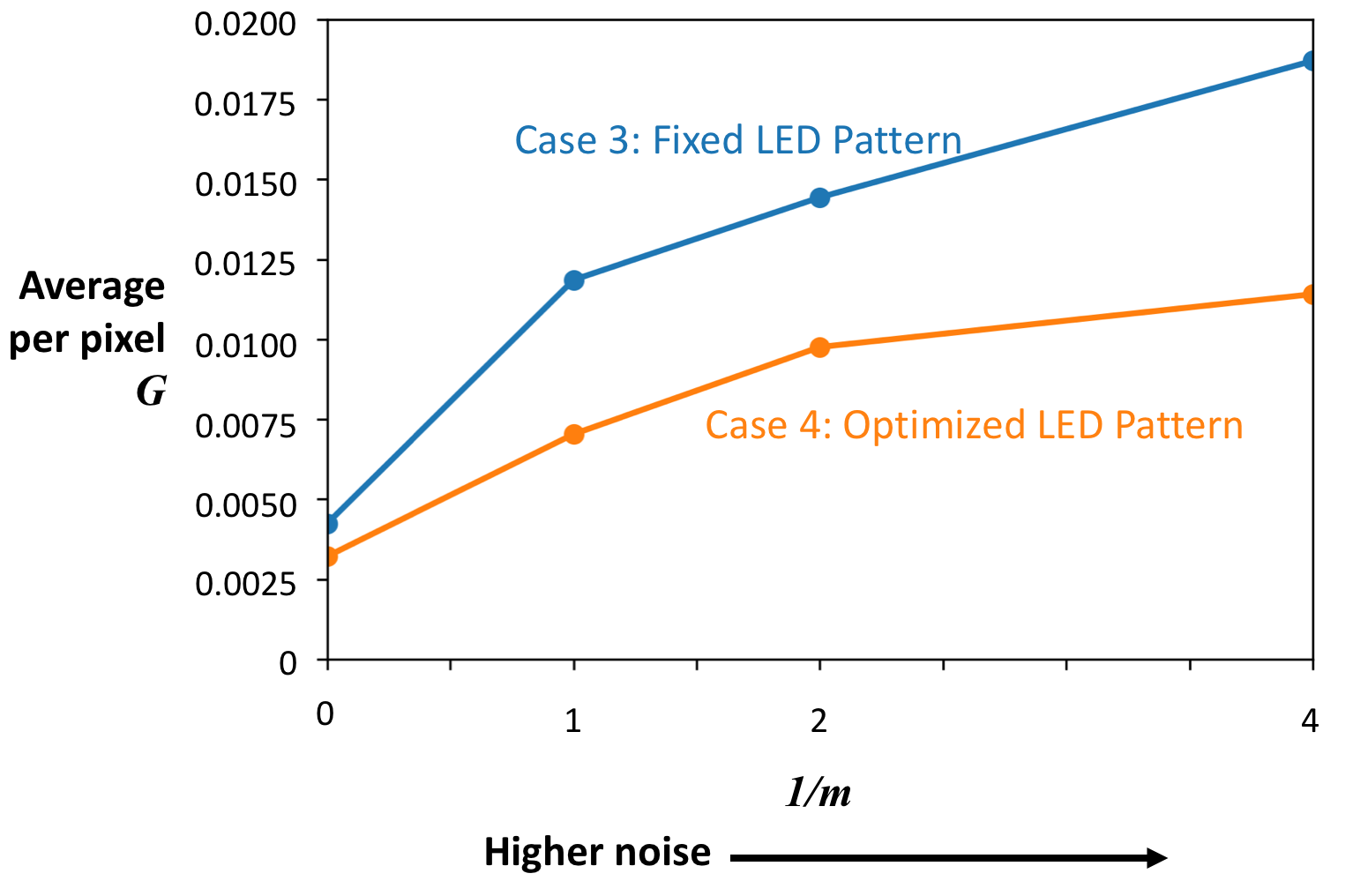}
  \caption{$G$ averaged over the test dataset for different $m$.}
  \label{fig:MNIST_Distinct_none_G_graph}
\end{subfigure}
\caption{Cases (3) and (4).}
\label{fig:MNIST_Distinct_none_graph}
\end{figure}

Fig.~\ref{fig:MNIST_Distinct_none_chart} shows the results for cases (3) and (4) for the same test example as in Fig.~\ref{fig:MNIST_Distinct_ioeo_chart}. We again plot $M$ and $G$ averaged over the entire test dataset for different levels of noise in Fig.~\ref{fig:MNIST_Distinct_none_graph} for cases (3) and (4). We note similar trends to cases (1) and (2): allowing the LED pattern to vary yields lower error. The final LED pattern in cases (2) and (4) are similar, even though the initial conditions were different. 

Though the MNIST dataset is useful for fast prototyping, the images do not resemble biological samples. In order to show proof-of-concept of this method for biologically relevant data, we utilized breast cancer cell images from the UCSB Bio-Segmentation Benchmark dataset \cite{bio-segmentation}, and converted the images to a high-resolution complex object dataset. Each image was cropped to 512 $\times$ 512 pixels and the three 8-bit color channels were summed. Each pixel $p_{0}$ of the resulting image was processed by:

\begin{equation}
e^{ \frac{i \pi \left(765 - p_{0} \right)}{765}},
\end{equation} 

\noindent and then low-pass filtered by the synthetic NA of the Fourier ptychographic microscope. We used 34 images from the UCSB Bio-Segmentation Benchmark dataset for training and 12 images each for validation and testing. The optical parameters used for this dataset of images are summarized in Table~\ref{table:BreastCancer_allLEDs}.

\begin{table}
\small
 \begin{center}
  \begin{tabular}{ | l | l |  }
    \hline
    Size of high-resolution object & 332.8 $\mu$m, 512 $\times$ 512 pixels \\ \hline 
    Wavelength of LED illumination & 630 nm \\ \hline 
    NA of microscope objective & 0.1 \\ \hline 
    Magnification of microscope objective & 5$\times$ \\ \hline 
    Pixel size of image sensor & 6.5 $\mu$m \\ \hline 
    Pitch of LED matrix & 4 mm \\ \hline 
    LED matrix size & 9 x 9 grid, center 69 LEDs allowed to vary \\ \hline 
    Height of LED matrix, $z_l$ & 100 mm \\ \hline 
    Synthetic NA of Fourier ptychographic microscope & 0.28 \\ \hline 
  \end{tabular}
\end{center}
\caption{Optical parameters used for the UCSB Bio-Segmentation Benchmark complex object dataset. \label{table:BreastCancer_allLEDs}}
\end{table}

\begin{figure}[htbp]
    \centering
    \includegraphics[scale=0.4]{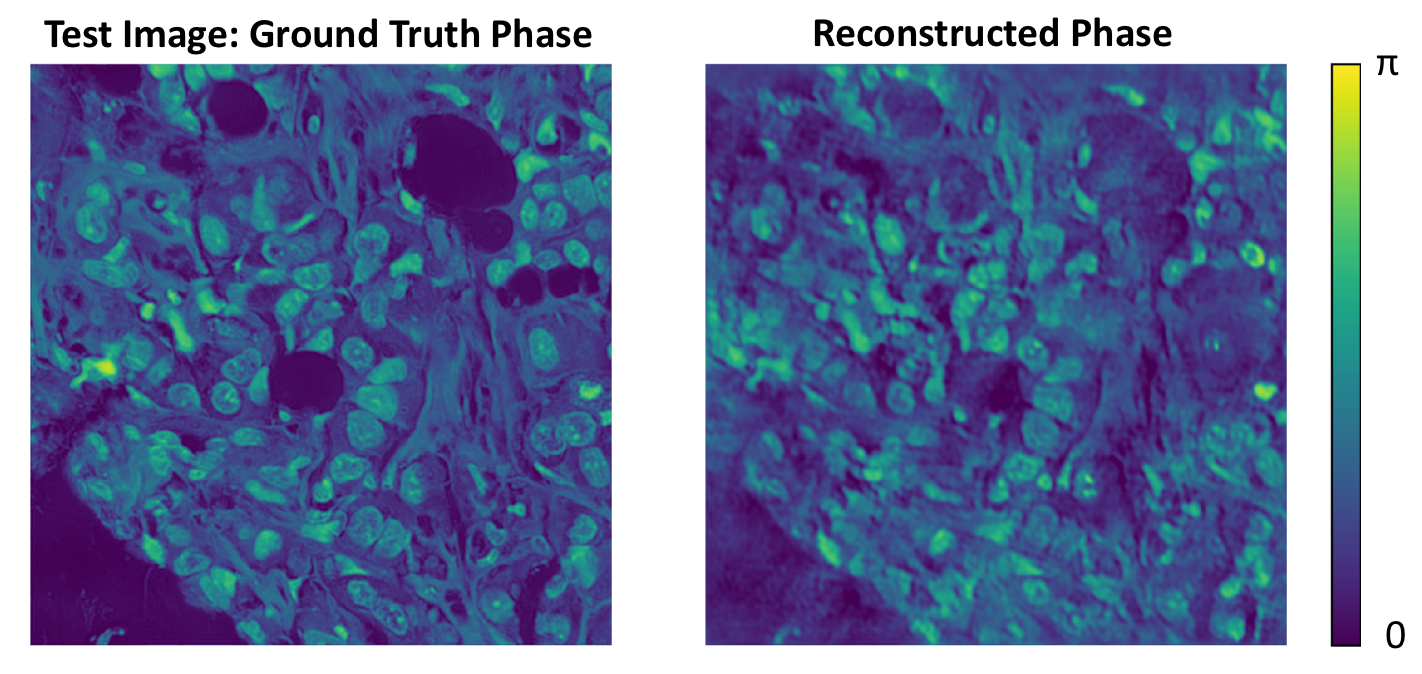}
    \caption{Test result for a single example for case (4) with the UCSB Bio-Segmentation Benchmark complex object dataset. The per pixel $M$ and $G$ for the reconstructed complex object are 0.112 and 0.014, respectively. Averaged over the entire training dataset, we have per pixel $M = 0.190$ and $G=0.016$.}
    \label{fig:BreastCancer_allLEDs_result}
\end{figure}

Fig.~\ref{fig:BreastCancer_allLEDs_result} shows a test example after training a neural network with noise factor $m=1$, following the considerations of case (4). We see reconstruction of many of the image details, and future work will focus on reducing the phase artifacts. 

\begin{figure}
\centering
\begin{subfigure}{1\textwidth}
  \centering
  \includegraphics[scale=0.5]{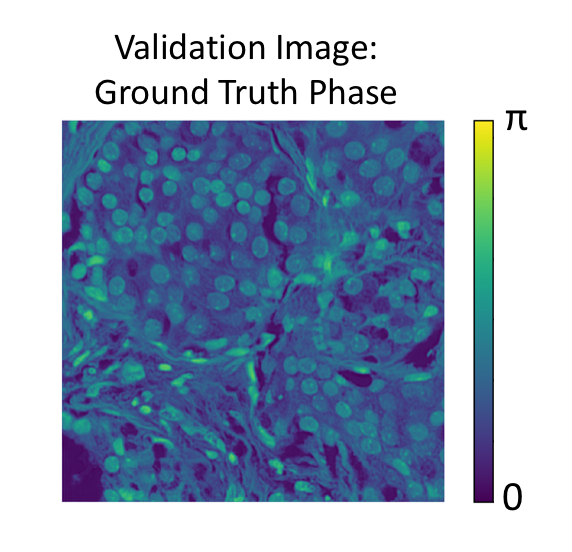}
  \caption{Example validation object phase.}
  \label{fig:BreastCancer_allLEDs_validation}
\end{subfigure}
\begin{subfigure}{1\textwidth}
  \centering
  \includegraphics[scale=0.45]{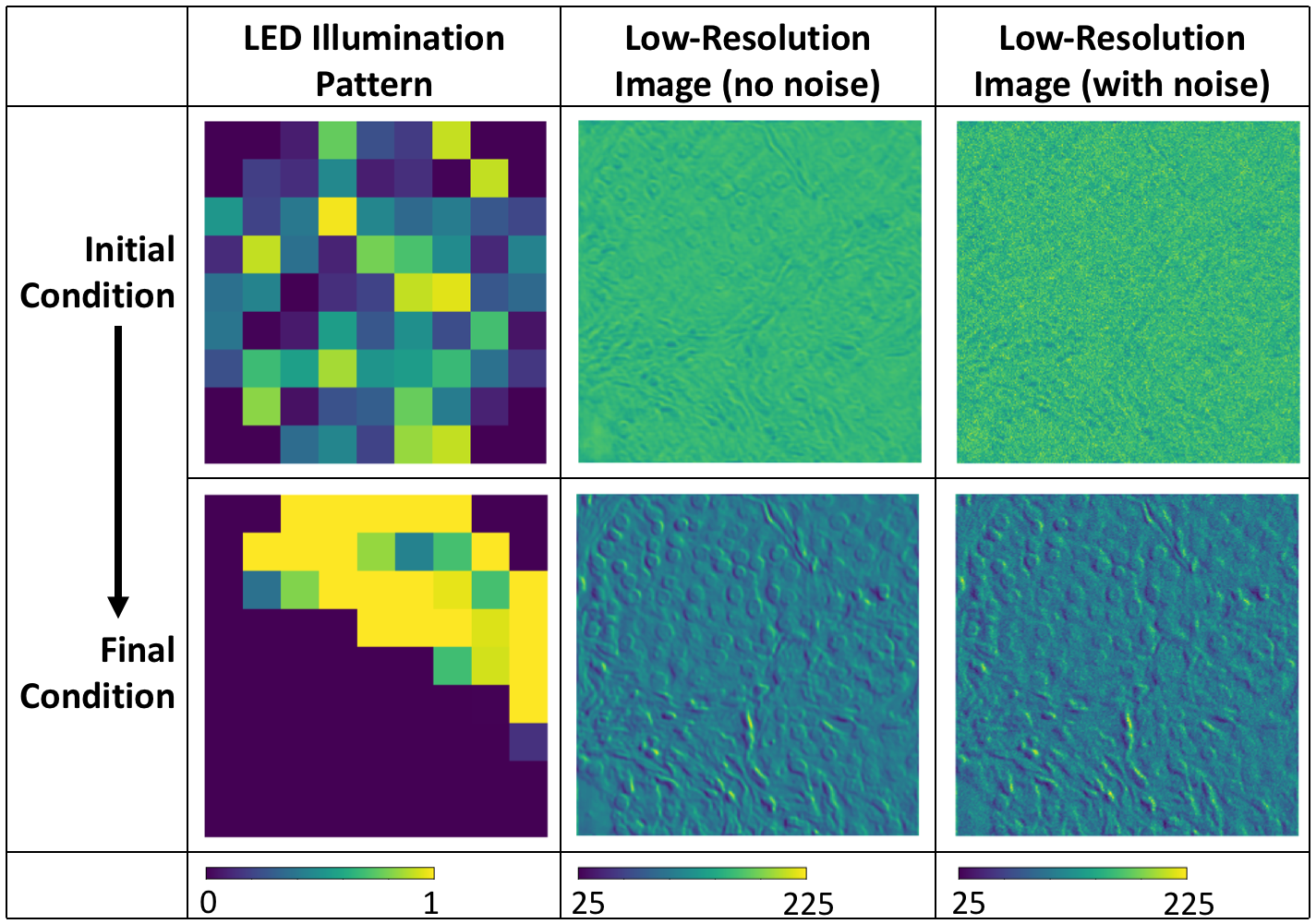}
  \caption{LED matrix and corresponding low-resolution images for the validation object, at the beginning and end of training.}
  \label{fig:BreastCancer_allLEDs_low_res}
\end{subfigure}
\label{fig:BreastCancer_allLEDs_low_res_val}
\end{figure}

Fig.~\ref{fig:BreastCancer_allLEDs_low_res} illustrates how the illumination pattern changed during training. The corresponding low-resolution images are also shown, with and without noise for the single validation example shown in Fig.~\ref{fig:BreastCancer_allLEDs_validation}. Visually, we see enhanced image contrast in the low-resolution image corresponding to the final illumination pattern. In the initial low-resolution image, the image contrast is almost totally obscured with the addition of noise. 

When we compare the low-resolution images with noise in Fig.~\ref{fig:BreastCancer_allLEDs_low_res}, it appears that more information about the original object is preserved the final image, corresponding to the optimized illumination pattern. We hypothesize that optimizing the illumination pattern increases the mutual information of the low-resolution image and high-resolution complex object. 

\begin{figure}[htbp]
    \centering
    \includegraphics[scale=0.4]{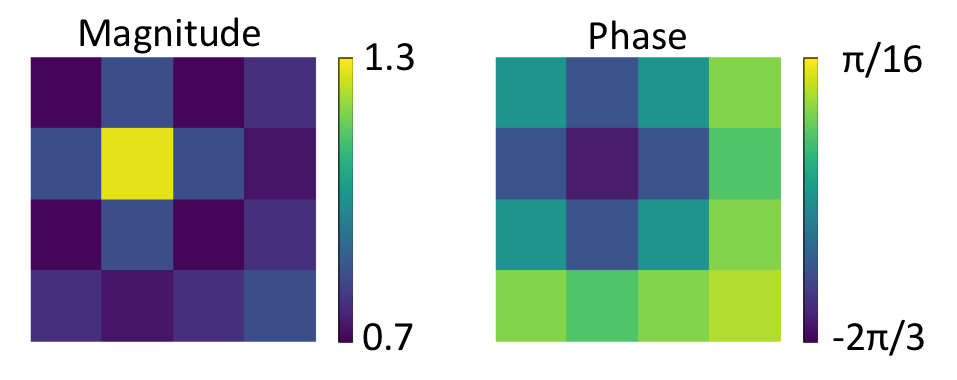}
    \caption{An example complex object in the 16 image, 4 $\times$ 4 pixel dataset.}
    \label{fig:Mutual_Info_16_example}
\end{figure}

We perform a simple experiment to determine if the mutual information increases as we train the deep neural network. We create a high-resolution dataset of 16 distinct images of 4 $\times$ 4 pixels. Each image is a pattern of ones and zeros and each pixel is processed by Eqn.~\ref{eq:mnist} and filtered by the synthetic NA of the Fourier ptychographic microscope. An example complex object of this dataset is shown in Fig.~\ref{fig:Mutual_Info_16_example}. The optical parameters used for this dataset are given in Table~\ref{table:Mutual_Info_16}. Given the optical parameters, each 4 $\times$ 4 pixel high-resolution complex object becomes a 1 $\times$ 1 pixel low-resolution intensity image. 

\begin{table}
\small
 \begin{center}
  \begin{tabular}{ | l | l |  }
    \hline
    Size of high-resolution object & 0.65 $\mu$m, 4 $\times$ 4 pixels \\ \hline 
    Wavelength of LED illumination & 630 nm \\ \hline 
    NA of microscope objective & 0.8 \\ \hline 
    Magnification of microscope objective & 10$\times$ \\ \hline 
    Pixel size of image sensor & 6.5 $\mu$m \\ \hline 
    Pitch of LED matrix & 8 mm \\ \hline 
    LED matrix size & 3 x 3 grid, all 9 LEDs allowed to vary \\ \hline 
    Height of LED matrix, $z_l$ & 10 mm \\ \hline 
    Synthetic NA of Fourier ptychographic microscope & 1.29 \\ \hline 
  \end{tabular}
\end{center}
\caption{Optical parameters for the 16 image, 4 $\times$ 4 pixel dataset. \label{table:Mutual_Info_16}}
\end{table}

\begin{figure}
\centering
\begin{subfigure}{.5\textwidth}
  \centering
  \includegraphics[scale=0.38]{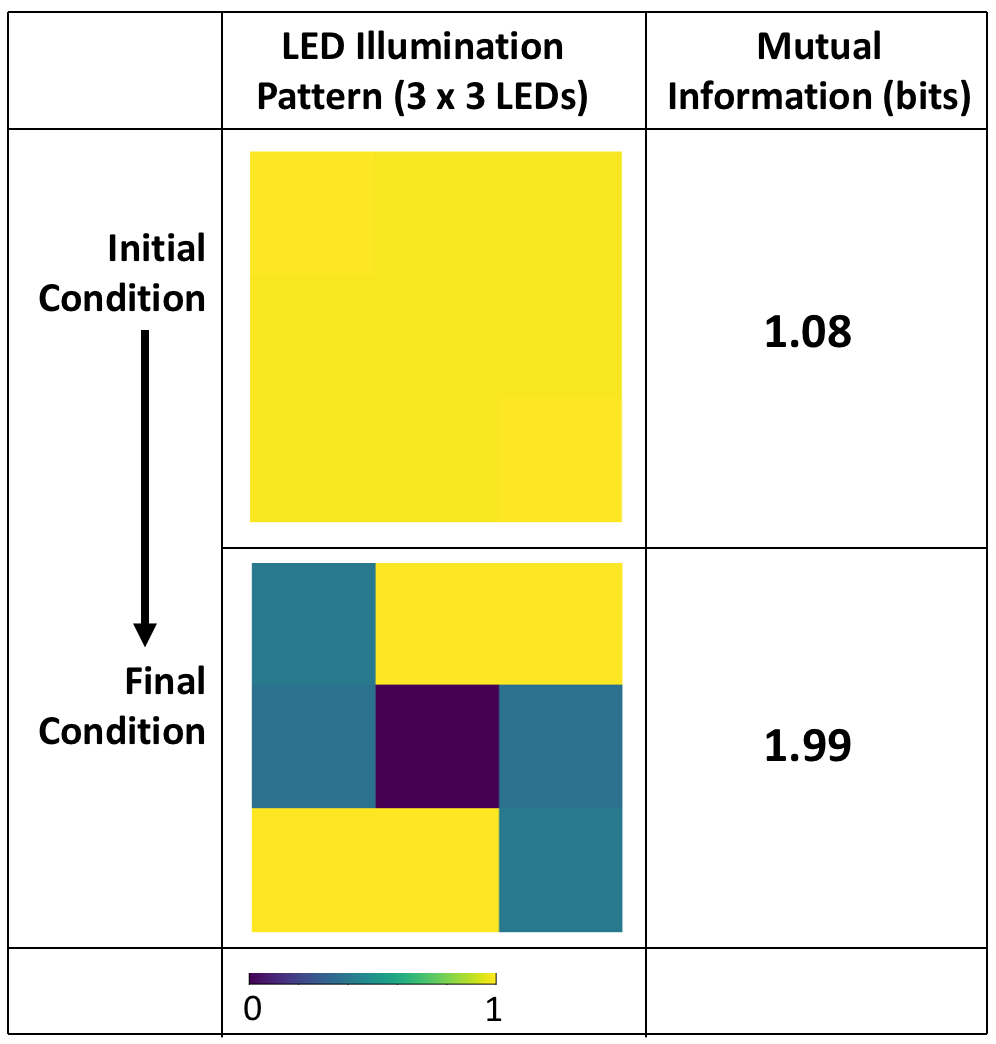}
  \caption{Case (2).}
  \label{fig:Mutual_Info_16_ioeo}
\end{subfigure}%
\begin{subfigure}{.5\textwidth}
  \centering
  \includegraphics[scale=0.38]{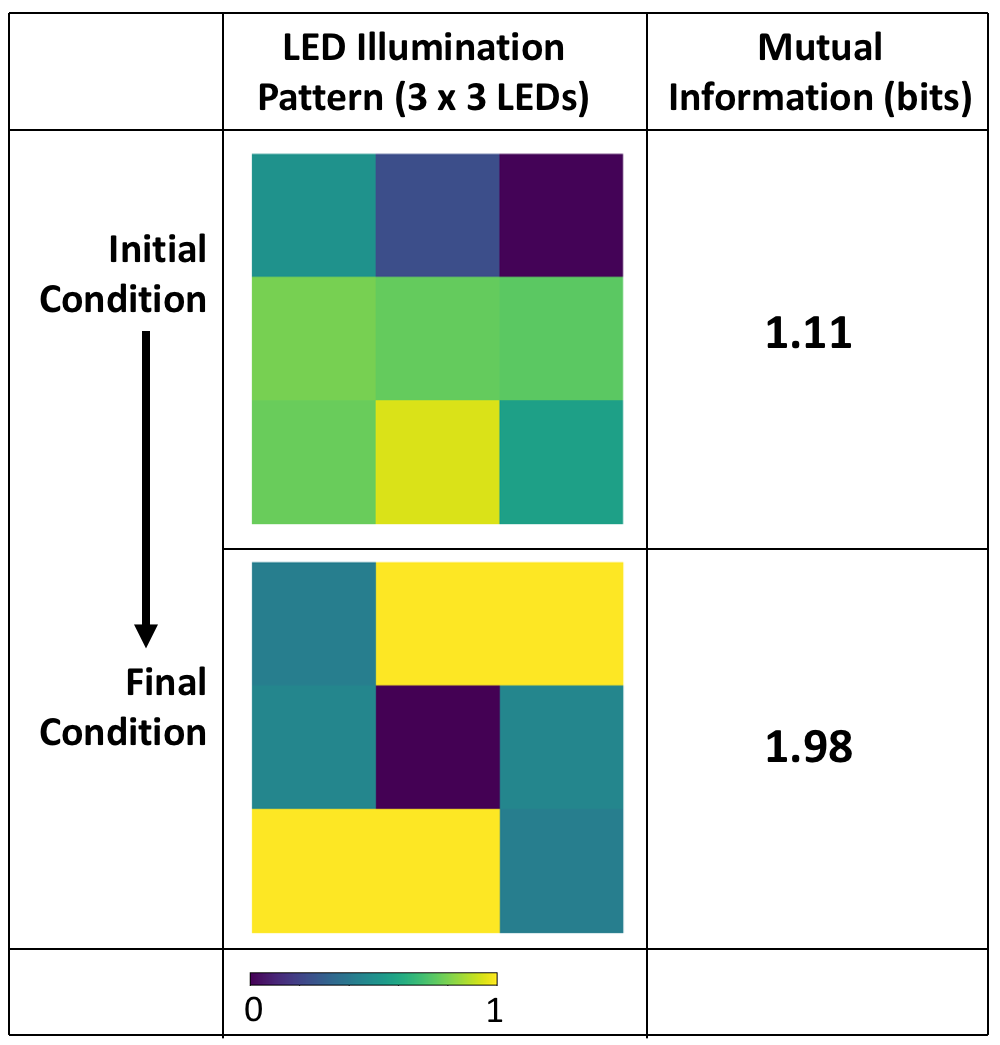}
  \caption{Case (4).}
  \label{fig:Mutual_Info_16_none}
\end{subfigure}
\caption{Mutual information of the high-resolution object and low-resolution image at the beginning and end of training.}
\label{fig:Mutual_Info_16}
\end{figure}

We trained 2 neural networks with no dropout, a batch size of 16, and noise level $m=1$, following the considerations of case (2) and case (4). The high-resolution dataset contains $\log_{2} {16} = 4$ bits of information. For the low-resolution image to capture all the information about the high-resolution dataset, the mutual information needs to be 4 bits. We calculate the mutual information for the initial and final illumination patterns of both neural networks, as shown in Fig.~\ref{fig:Mutual_Info_16}. The mutual information is calculated by:

\begin{equation}
\sum_{y \in Y} \sum_{x \in X} p \left( x,y \right) \log_2 \left( \frac{p \left( x,y \right)}{ p(x) p(y) } \right), 
\end{equation}

\noindent where $p(x)$ is the marginal probability distribution of high-resolution objects, $p(y)$ is the marginal probability distribution of low-resolution images, and $p(x,y)$ is the joint probability distribution. For every $x \in X$, we use Eqn.~\ref{eq:noise} to generate 1,000,000 samples of $y$ and approximate the distribution $p(y \mid X = x)$.

From Fig.~\ref{fig:Mutual_Info_16}, we see that in our joint optimization procedure of the illumination pattern and the post-processing parameters, the mutual information of the low-resolution image and high-resolution object increases. We also note that the final LED illumination patterns are similar for cases (2) and (4). It appears that during training, the LED illumination pattern changes so that more information about the high-resolution object is physically encoded in the low-resolution image. The physical preprocessing step is crucial to improved final complex object reconstruction.

\section{Conclusions}

The success of deep learning presents an opportunity in the design of imaging systems: by finding generalizable structure in past data, it is possible to figure out how to optimally collect future data. This work uses deep learning to jointly optimize the physical parameters of the imaging system together with the parameters of the image post-processing algorithm. We consider the optimization of Fourier ptychographic microscopy and demonstrate that we can eliminate the tradeoff among spatial resolution, field-of-view, and temporal resolution for a given sample type.

This work frames the task of imaging a high-resolution complex field as a communications problem, where image information is: (1) encoded as an optical signal with the programmable illumination source of a Fourier ptychographic microscope, (2) collected by an image sensor, and (3) computationally decoded. With a training dataset of high-resolution phase images of fixed cells, the parameters of both the microscope and the decoding algorithm can be co-optimized to maximize imaging speed. The optimized microscope can then be implemented to image the live sample, and the collected measurements fed directly into the post-processing layers of the computational graph, allowing for a high single-shot space-bandwidth product. With this method, instead of collecting images in the traditional sense, we optimally collect information about the spatial structure of the live sample. By using prior knowledge gained from static images, we avoid redundancy in information collection by physically preprocessing the data.

We present results from simulated data, in order to show the feasibility of this method. Future work is needed to validate these methods with real experimental data. Additionally, our work assumes that the probability distribution of the training data matches the probability distribution of the test data. It is possible that the structure of the fixed samples may not be the same as the live samples. A possible method to alleviate this problem is to feed the output of the deep neural network into an iterative solver to ensure a valid solution to the inverse problem. Finally, we do not consider temporal relationships between images in this work. We consider each image frame as independent, however there is clearly a relationship between sequential frames. With a training set of video sequences, these temporal relationships could be accounted for. 

The framework outlined in this paper is not specific to Fourier ptychographic microscopy; these ideas can be extended to other imaging modalities with programmable parameters. Though we have focused on the benefits of this method for imaging of live biological samples, other benefits include reduced data storage needs and faster imaging for high-throughput studies. 

\section*{Acknowledgments}

This work used the Extreme Science and Engineering Discovery Environment (XSEDE) XStream at Stanford University through allocation TG-TRA100004. XSEDE is supported by National Science Foundation grant number ACI-1548562. 

The authors would like to thank Eden Rephaeli for coining the term ``physical preprocessing."

\bibliography{FPM_Paper}

\end{document}